\documentclass[lettersize,journal]{IEEEtran}
\usepackage{amsmath,amsfonts}
\usepackage{algorithmic}
\usepackage{algorithm}
\usepackage{array}
\usepackage{subfig}
\usepackage{textcomp}
\usepackage{stfloats}
\usepackage{url}
\usepackage{verbatim}
\usepackage{graphicx}
\usepackage{cite}
\usepackage{hyperref}
\hypersetup{
  colorlinks=true,
  linkcolor=blue,
  citecolor=blue,
  urlcolor=blue
}

\begin{document}

\title{NN-Based Joint Mitigation of IQ Imbalance and PA Nonlinearity With Multiple States}

\author{Yundi Zhang, Wendong Cheng and Li Chen,~\IEEEmembership{Senior Member,~IEEE}
        % <-this % stops a space
\thanks{This work was supported by the Natural Science Foundation of China (Grant No. 62471449) and Anhui Provincial Natural Science Foundation (No.2308085J24). (Corresponding author: Li Chen.)}% <-this % stops a space
\thanks{Yundi Zhang, Wendong Cheng and Li Chen are with the CAS Key Laboratory of Wireless Optical Communication, University of Science and Technology of China (USTC), Hefei 230027, China (e-mail: zyd027570@mail\\.ustc.edu.cn; cwd01@mail.ustc.edu.cn; chenli87@ustc.edu.cn).}}

% The paper headers
%\markboth{Journal of \LaTeX\ Class Files,~Vol.~14, No.~8, August~2021}%
%{Shell \MakeLowercase{\textit{et al.}}: A Sample Article Using IEEEtran.cls for IEEE Journals}

%\IEEEpubid{0000--0000/00\$00.00~\copyright~2021 IEEE}
% Remember, if you use this you must call \IEEEpubidadjcol in the second
% column for its text to clear the IEEEpubid mark.

\maketitle

\begin{abstract}
Joint mitigation of IQ imbalance and PA nonlinearity is important for improving the performance of radio frequency (RF) transmitters. 
In this paper, we propose a new neural network (NN) model, which can be used for joint digital pre-distortion (DPD) of 
non-ideal IQ modulators and PAs in a transmitter with multiple operating states. 
The model is based on the methodology of multi-task learning (MTL). 
In this model, the hidden layers of the main NN are shared by all signal states, 
and the output layer's weights and biases are dynamically generated by another NN. 
The experimental results show that the proposed model can effectively perform joint DPD for IQ-PA systems, and it 
achieves better overall performance within multiple signal states than the existing methods. 
\end{abstract}


\section{Introduction} 
\IEEEPARstart{I}n radio frequency (RF) transmitters, IQ imbalance and PA nonlinearity are two major hardware impairments. 
PA nonlinearity is caused by operating at an efficient point, and it induces harmonics, intermodulation distortion, and spectral regrowth. 
IQ imbalance is caused by the gain and phase mismatch, 
which is the result of imperfect local oscillators (LOs), 
and the nonideal impulse response of the analog devices in the I and Q branches, 
and it can degrade the modulated signal's quality. 
Therefore, mitigating these hardware impairments is very important for improving the performance of RF transmitters. 

Various methods have been proposed to mitigate them separately. 
For PA nonlinearity, the most common solution is digital predistortion (DPD). 
In \cite{b1}, researchers proposed a memory polynomial (MP) model for DPD.  
The authors in \cite{b2} improved the MP model and proposed a generalized memory polynomial (GMP) model. 
To mitigate IQ imbalance, the authors in \cite{b3} proposed a training sequence based method. 
Researchers in \cite{b4} proposed a blind method that didn't require sending training sequences in advance. 

To mitigate PA nonlinearity and IQ imbalance jointly, some more powerful DPD models have been proposed, and most of them were based on neural networks (NNs). 
In \cite{b5}, researchers proposed a real-valued focused time-delay NN (RVFTDNN) model based on the multilayer perceptron (MLP) architecture. 
Further, on the basis of the RVFTDNN, researchers in \cite{b6} and \cite{b8} proposed the augmented real-valued time-delay NN (ARVTDNN) and the shortcut real-valued time-delay NN (SVDEN), respectively. 

Most of the existing joint DPD models (i.e., DPD models that can jointly mitigate PA nonlinearity and IQ imbalance), 
including those mentioned above, were only applicable when the signal states (e.g., bandwidth and power of the input signal) were stable. 
But in modern wireless communication systems, the signal states change dynamically\cite{b14}. 
So it's important to design a joint DPD model that can perform well within multiple states. 

Note that, when only considering PA nonlinearity, some studies on DPD have taken multiple signal states into account. 
The authors in \cite{b9} proposed a DPD model that can adapt to changes in temperature and power. 
In \cite{b12}, researchers proposed a gated dynamic NN (GDNN) model to linearize PAs with multiple states. 
However, these works didn't consider IQ imbalance, and these methods' performance is limited when both IQ imbalance and PA nonlinearity are present. 
To the best of our knowledge, the joint mitigation of IQ imbalance and PA nonlinearity with multiple states has never been discussed.

In this work, we propose a new NN-based joint DPD model that can achieve excellent performance within multiple states. 
The model is based on the “hard parameter sharing” in multi-task learning (MTL). 
The hidden layers of the main NN are shared by all signal states. 
At the same time, we use another NN to generate different weights and biases of the main NN's output layer for each signal state separately. 
Experimental results show that, the proposed model can achieve excellent performance in terms of normalized mean square error (NMSE) and adjacent channel power ratio (ACPR), 
and it needn't be retrained when the states change. 

\section{Preliminaries}\label{S2}
\subsection{IQ Imbalance and PA Nonlinearity}
\begin{figure}[t]
    \centering
    \includegraphics[width=1\linewidth]{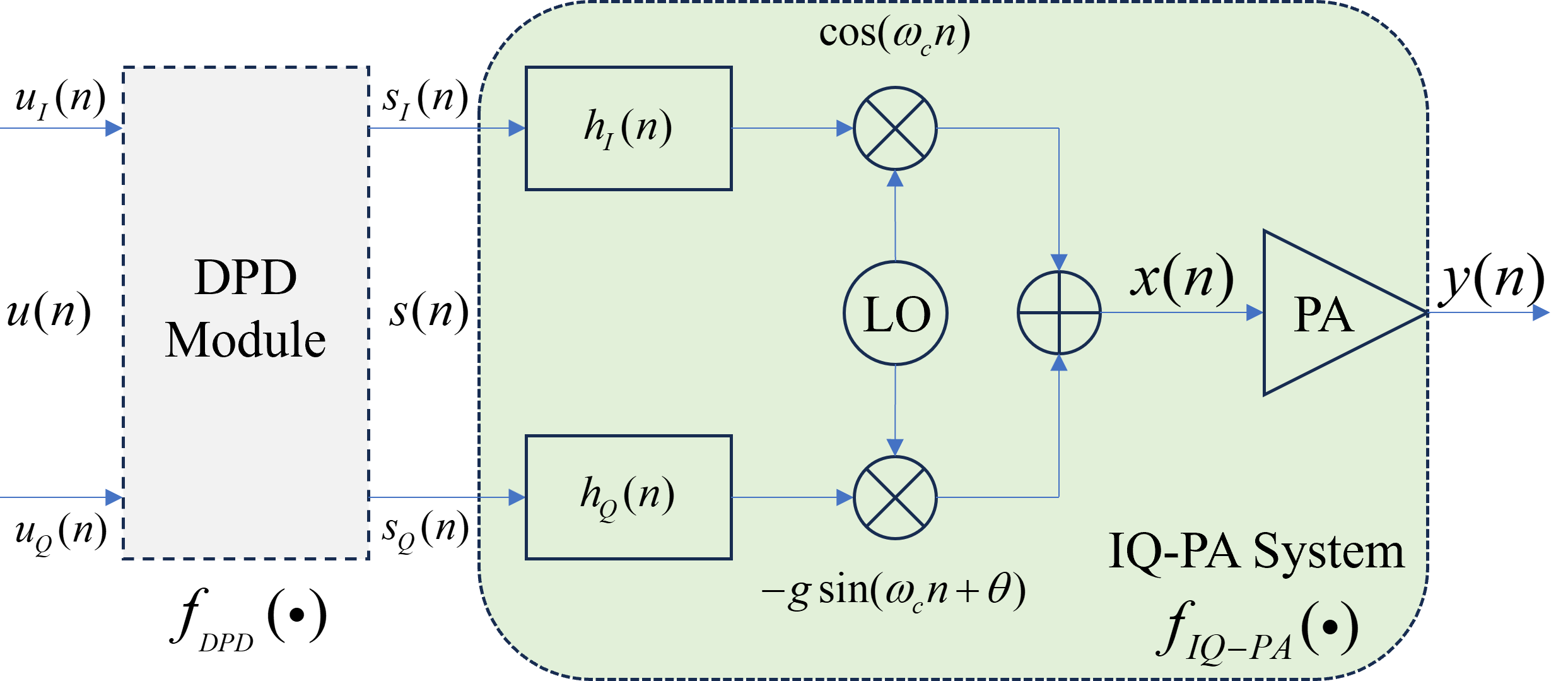}
    \caption{Complex baseband equivalent model of the IQ-PA system and the basic idea of joint DPD}
    \label{fig1}
\end{figure}

As shown in Fig.~\ref{fig1}, we consider the complex baseband equivalent model of the IQ-PA system.
$s(n)$ is the input baseband signal, $s_\text{I}(n)$ and $s_\text{Q}(n)$ are its real and imaginary parts.
$h_\text{I}(n)$ and $h_\text{Q}(n)$ are the equivalent impulse responses of devices (e.g., DACs, LPFs) 
in the I and Q branches. $g$ and $\theta $ are the gain and phase mismatch. 
$x(n)$ and $y(n)$ are the output signal of the IQ modulator and the PA, respectively. 

The signal upconverted by the non-ideal IQ modulator is
\begin{equation}\label{eq1}
    x\left( n \right) =s\left( n \right) \ast K_\text{1}\left( n \right) +s^*\left( n \right) \ast K_\text{2}\left( n \right),
\end{equation}
where $\ast $ denotes the convolution operation and
\begin{equation}\label{eq2}
    \begin{split}
        &K_\text{1}\left( n \right) =\frac{1}{2}\left[ h_\text{I}\left( n \right) +ge^{j\theta}h_\text{Q}\left( n \right) \right]\\
        &K_\text{2}\left( n \right) =\frac{1}{2}\left[ h_\text{I}\left( n \right) -ge^{j\theta}h_\text{Q}\left( n \right) \right].
    \end{split}
\end{equation}
There, $g$ and $\theta $ reflect the frequency-independent IQ imbalance, $h_\text{I}(n)$ and $h_\text{Q}(n)$ reflect the 
frequency-dependent IQ imbalance.
Ideally, $h_\text{I}\left( n \right) =h_\text{Q}\left( n \right) =\delta \left( n \right)$, where
$\delta \left( n \right)$ denotes the unit impulse response, and $g=1$, $\theta =0$.
Substituting into \eqref{eq1} we have $x(n)=s(n)$.

When operating under large signal bandwidth, PAs will exhibit significant memory effects, i.e., the output signal of the PA at the current moment is related not only to the input at the current moment, but also to the inputs at several previous moments. 
So the equivalent complex baseband signal of the PA output radio frequency (RF) signal can be expressed as
\begin{equation}\label{eq3}
    y\left( n \right) =f_\text{PA}\left[ x\left( n \right) ,x\left( n-1 \right) ,\ldots  ,x\left( n-M_\text{PA} \right) \right],
\end{equation}
where $f_\text{PA}(\cdot )$ is a function representing the PA and $M_\text{PA}$ denotes its memory length.

In order to jointly consider IQ imbalance and PA nonlinearity, 
we combine \eqref{eq1} and \eqref{eq3} as
\begin{equation}\label{eq4}
    y\left( n \right) =f_\text{IQ-PA}\left[ s\left( n \right) ,s\left( n-1 \right) ,\ldots ,s\left( n-M_\text{IQ-PA} \right) \right],
\end{equation}
where $f_\text{IQ-PA}(\cdot )$ is a function representing the IQ-PA system and $M_\text{IQ-PA}$ denotes its total memory length.
It can be seen that, due to the coupling of the IQ imbalance and the PA nonlinearity, 
it's difficult to express the input-output characteristics of the IQ-PA system in detail.

When operating states (e.g., bandwidth and power states of the signal) change, the IQ modulator and PA exhibit different characteristics. 
Assume that there are $L$ types of different states, the superscript $i\in \left\{ 1,2,\ldots ,L \right\}$ indicates one specific state, and 
\begin{equation}\label{eq5}
    y^{\left( i \right)}\left( n \right) =f_\text{IQ-PA}^{\left( i \right)}\left[ s^{\left( i \right)}\left( n \right) ,\ldots ,s^{\left( i \right)}\left( n-M_\text{IQ-PA} \right) \right].
\end{equation}
Note that the characteristics of $f_\text{IQ-PA}^{\left( i \right)}(\cdot)$ change when $i$ takes different values. 
Therefore, it will be even more difficult to express the IQ-PA system in detail with multiple states.

\subsection{The Problem of Joint DPD}\label{S2-B}
As shown in Fig.~\ref{fig1}, by cascading a module with opposite nonlinear characteristics in front of the IQ-PA system in the digital domain, the DPD-IQ-PA system can exhibit nearly ideal characteristics\cite{b5}\cite{b6}\cite{b8}, i.e., 
$\tilde{y}\left( n \right) \approx u\left( n \right)$, where $\tilde{y}\left( n \right)$ represents the output of the IQ-PA system after deploying the DPD module, and $u(n)$ is the input of the DPD module. 
$\tilde{y}\left( n \right)$ can be expressed as 
\begin{equation}\label{eq6}
    \tilde{y}(n)\!=\!f_{\text{IQ-PA}}\big[f_{\text{DPD}}\big(\boldsymbol{u}(n)\big),\ldots,f_{\text{DPD}}\big(\boldsymbol{u}(n\!-\!M_{\text{IQ-PA}})\big)\big],
\end{equation}
where $f_\text{DPD}(\cdot )$ is a function representing the selected DPD model, $\boldsymbol{u}\left( n \right) =\left[ u\left( n \right) ,\ldots ,u\left( n-M_\text{DPD} \right) \right]$ 
is its input vector and $M_\text{DPD}$ is its memory length. 

When there are multiple operating states, equation \eqref{eq6} can be rewritten as 
\begin{equation}\label{eq7}
    \tilde{y}^{(i)}\!(n)\!=\!f_{\text{IQ-PA}}^{(i)}\!\big[f_{\text{DPD}}\!\big(\!\boldsymbol{u}^{\!(i)}\!(n)\!\big),\ldots,f_{\text{DPD}}\!\big(\!\boldsymbol{u}^{\!(i)}\!(n\!-\!M_{\text{IQ-PA}})\!\big)\!\big],
\end{equation}
where $i\in \left\{ 1,2,\ldots ,L \right\}$. It can be seen that, 
in order to achieve excellent calibration performance within all states, $\tilde{y}^{(i)}(n) \approx u^{(i)}(n)$ must hold for every value of $i$, 
and the characteristics of $f_\text{IQ-PA}^{(i)}(\cdot )$ and $f_\text{DPD}(\cdot )$ must match each other within all states, which is much more challenging.

\subsection{Existing Methods}
\begin{figure}[t]
   \centering
   \subfloat[]{\label{fig2-a}
   \includegraphics[width=0.95\linewidth]{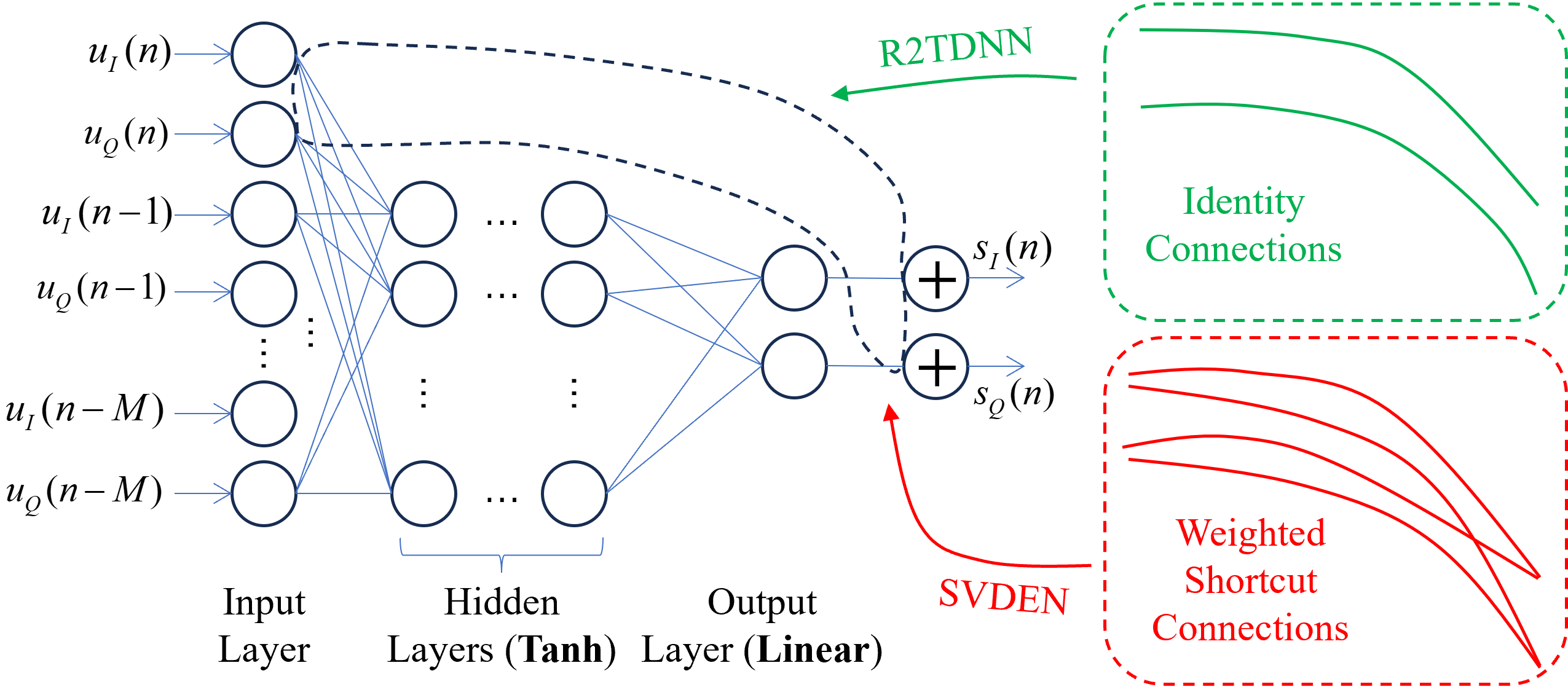}}

   \subfloat[]{\label{fig2-b}
   \includegraphics[width=0.95\linewidth]{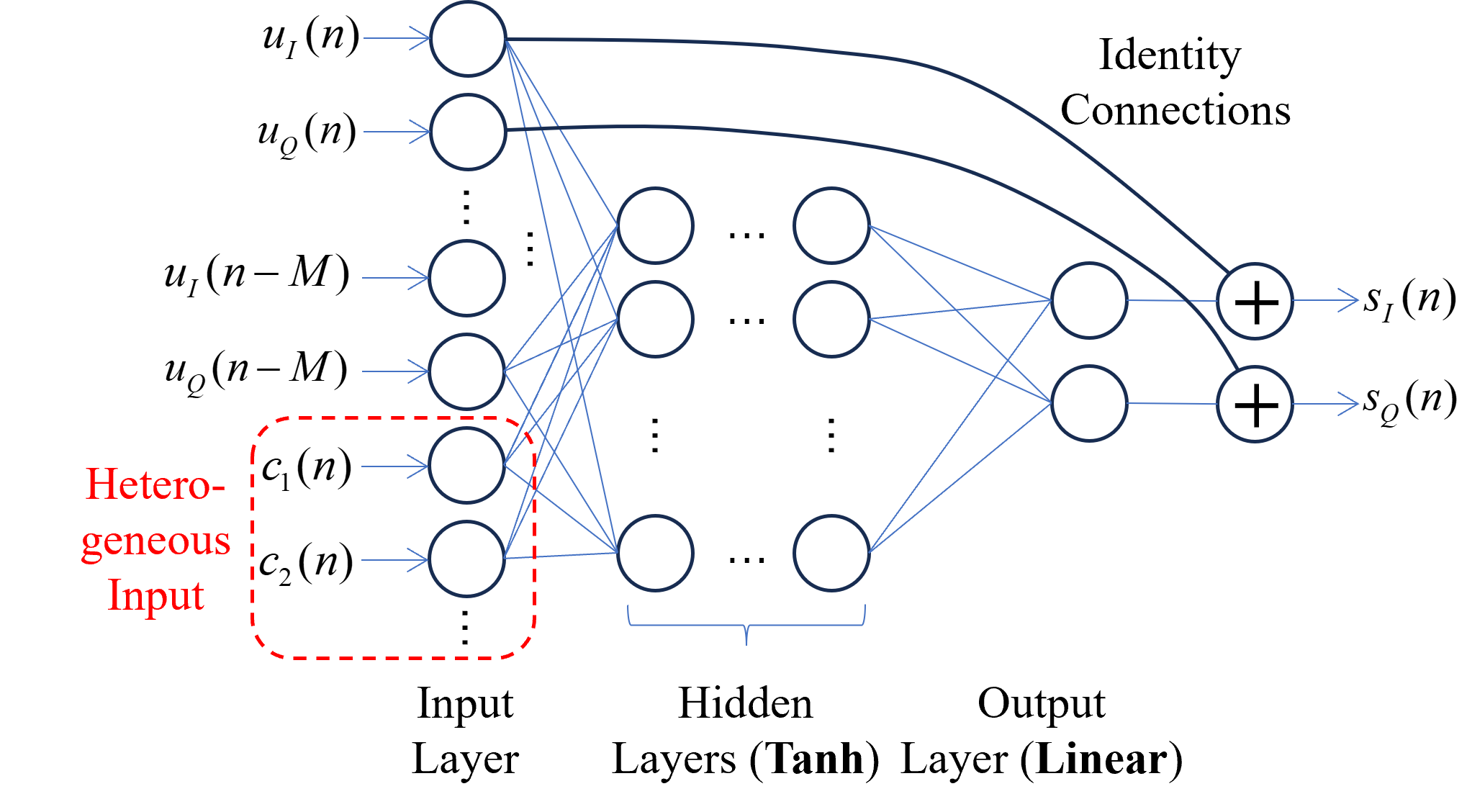}}

   \caption{Existing models. (a) R2TDNN and SVDEN. (b) HG-R2TDNN}
   
    \label{fig2}
\end{figure}
Next, we will briefly introduce three existing methods. 

\begin{itemize}
  \item \textbf{R2TDNN\cite{b7}}: R2TDNN can be used for IQ-PA joint DPD with fixed states, and its block diagram is shown in Fig.~\ref{fig2}\subref{fig2-a}. 
  It receives the input baseband signal's real and imaginary parts at the current moment and several past moments, and outputs the pre-distorted signal at the current moment. 
  Moreover, the two neurons fed by the current input signal are connected to the output neurons by the identity connections, so the model 
  only needs to learn the difference between the ideal mapping and the identity mapping. 
  \item \textbf{SVDEN\cite{b8}}: SVDEN was also designed for joint DPD with fixed states. 
  As shown in Fig.~\ref{fig2}\subref{fig2-a}, compared to R2TDNN, SVDEN uses 4 weighted shortcut connections to replace the 2 identity connections used by R2TDNN. 
  So SVDEN can more flexibly learn the linear part of the ideal mapping. 
  \item \textbf{Heterogeneous NN\cite{b9}\cite{b10}}: 
  When there are multiple operating states, adding heterogeneous inputs to the model is a common strategy. 
  As shown in Fig.~\ref{fig2}\subref{fig2-b}, 
  by directly adding operating states to the input layer of the NN (these inputs are often referred to as the heterogeneous inputs), 
  the DPD model can adjust its characteristics according to these inputs and dynamically match the system to be calibrated. 
  In this paper, we choose the R2TDNN as its main part, so it is called HG-R2TDNN. 
\end{itemize}

It's worth pointing out that, the R2TDNN and the SVDEN's performance will degrade drastically when operating states change, 
and the heterogeneous NN's performance is limited when both IQ imbalance and PA nonlinearity are considered. 
Next, we'll introduce the proposed model, which can effectively perform joint DPD for IQ-PA systems with multiple states. 

\section{Proposed Joint DPD Model}\label{S3}
\subsection{The Proposed HN-R2TDNN}
The proposed model is based on the idea of multi-task learning (MTL). 
MTL aims to jointly learn multiple related tasks so that the knowledge in one task can be exploited by others, and it is expected to improve 
the generalization ability and performance of the model on all tasks\cite{b19}. 
It has already been found that PAs behave similarly within different states\cite{b18}, so the methodology 
of MTL is expected to be useful in dynamic joint DPD (i.e., joint DPD with multiple states). 

We use the approach of hard parameter sharing (HPS) \cite{b13} to perform NN-based MTL. 
The basic idea of HPS is to share the hidden layers for all tasks and design different output layers for each specific task. 
However, there are two key points to consider when designing the joint DPD model. First, designing a separate output layer for each state would significantly increase the overall complexity of the model. 
Second, the common HPS can only be applied when the operating states are discrete and finite. 

\begin{figure}[t]
    \centering
    \subfloat[]{\label{fig3-a}
    \includegraphics[width=0.95\linewidth]{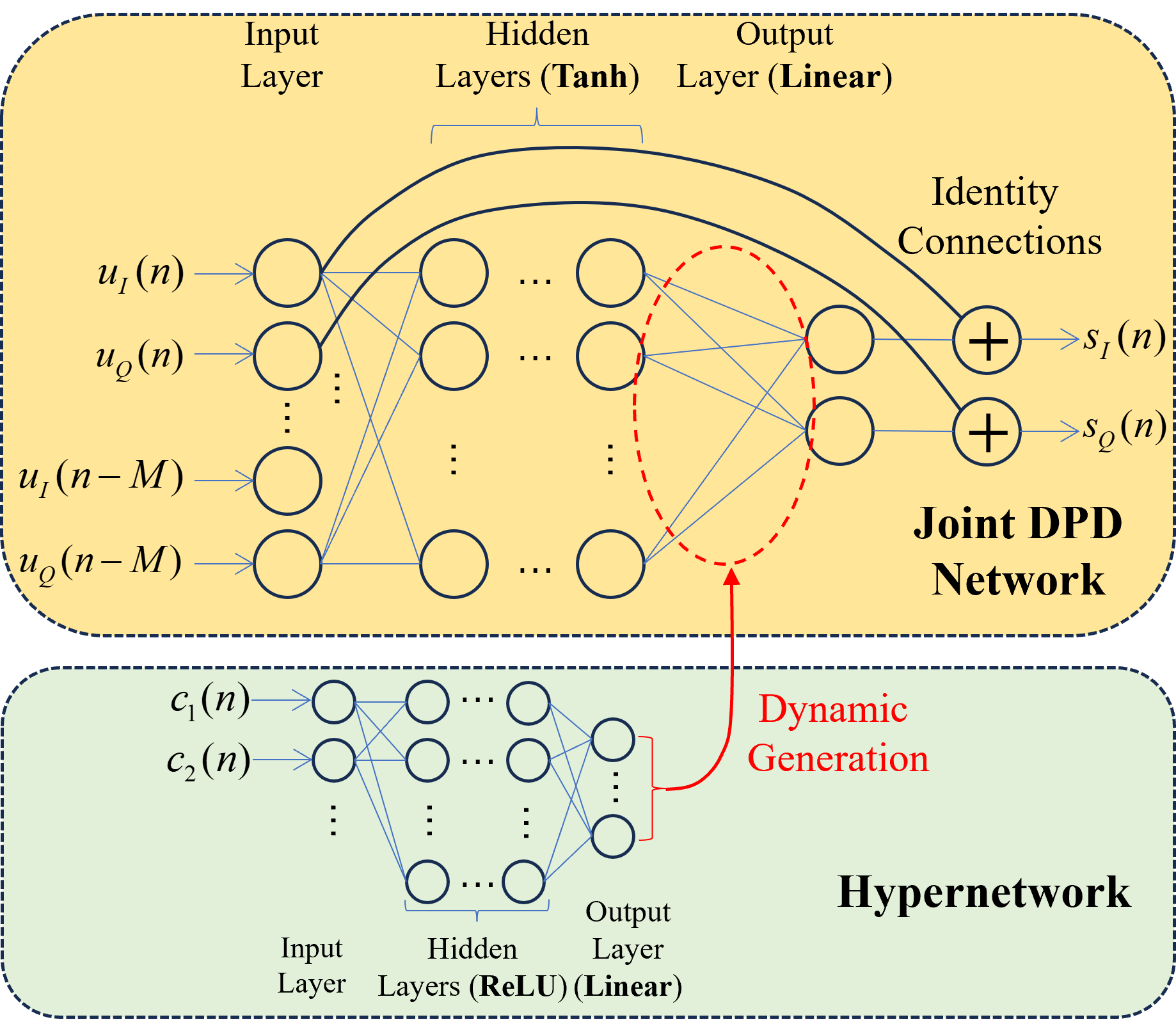}}

    \subfloat[]{\label{fig3-b}
    \includegraphics[width=0.95\linewidth]{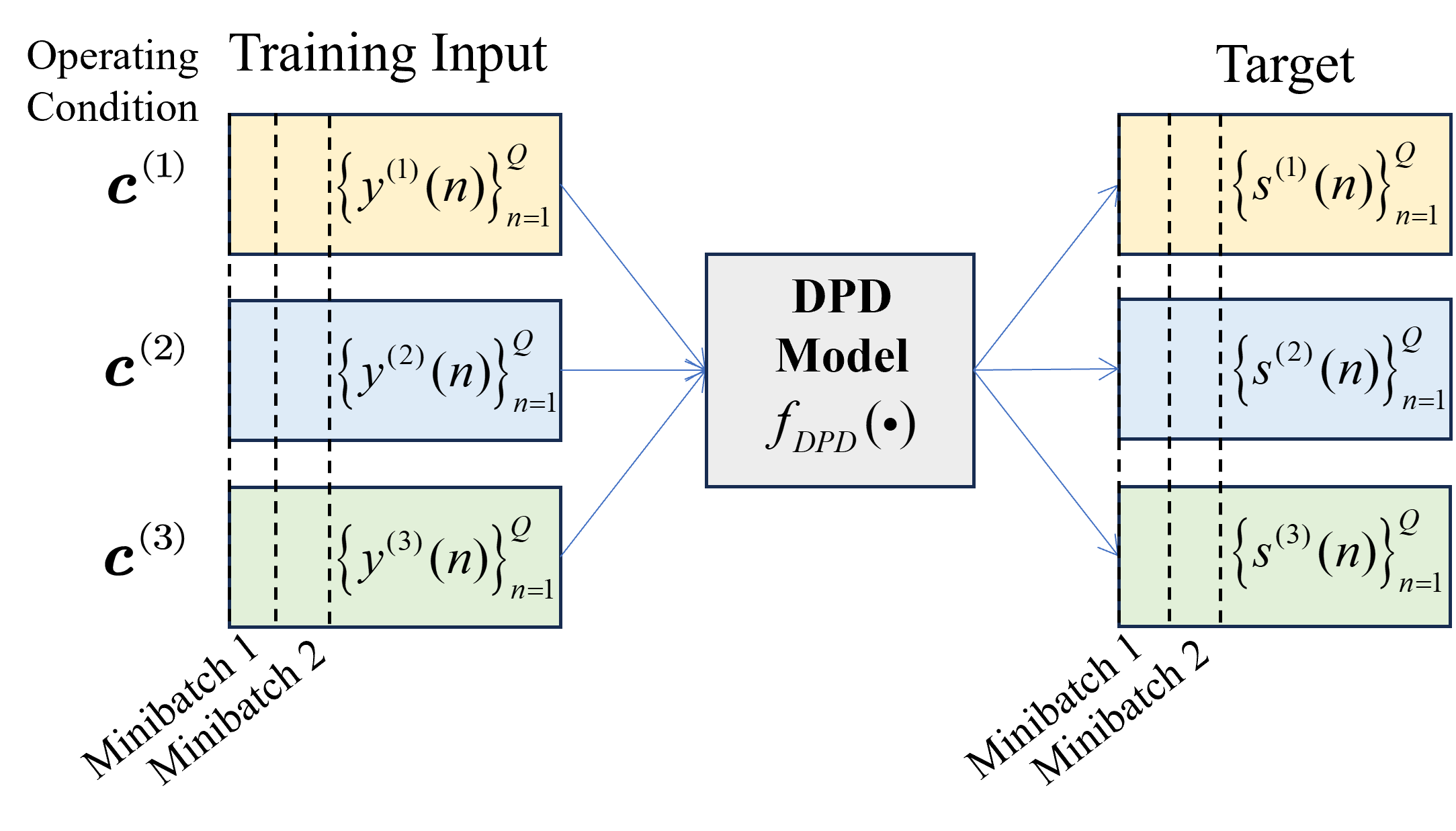}}

    \caption{Proposed Method. (a) Proposed HN-R2TDNN for joint DPD with multiple states. (b) The parallel loading approach of training data when using ILA}
    \label{fig3}
\end{figure}

Considering the above two key points, we propose the HN-R2TDNN, as shown in Fig.~\ref{fig3}\subref{fig3-a}. 
The model consists of two parts. The first part is the joint DPD network, 
which receives the input baseband signal at the current moment and several past moments, 
and outputs the pre-distorted signal at the current moment. 
We choose the R2TDNN as this part. 
The second part is called “hypernetwork”, 
which receives the operating states at the current moment and dynamically generates weights and biases for the joint DPD network's output layer. 
We choose the multilayer perceptron (MLP) architecture to build this part, 
where the activation function of the hidden layers is the rectified linear unit (ReLU) and the output layer is a linear layer with no activation function. 

The details of the proposed HN-R2TDNN can be described as follows. 
Assume that the joint DPD network has a total of $R$ layers, where there are $R-2$ hidden layers, $1$ input layer and $1$ output layer. 
$D_{r}$ denotes the number of neurons in layer $r$, and $D_{1}=2M+2$, $D_{R}=2$, where $M$ is the memory length of the model. 
$\boldsymbol{W}_r\in \mathbb{R}^{D_r\times D_{r-1}}$ is the weight matrix that connects layer $r-1$ and $r$, 
$\boldsymbol{b}_r\in \mathbb{R}^{D_r}$ and $\boldsymbol{u}_r$ are vectors representing the biases and outputs (or activations) of layer $r$, respectively, 
and $2\leq  r\leq  R$. 
The input vector of the joint DPD network is denoted as 
\begin{equation}\label{eq8}
    \boldsymbol{u}_1=\left[ u_\text{I}\left( n \right) ,u_\text{Q}\left( n \right) ,\ldots ,u_\text{I}\left( n-M \right) ,u_\text{Q}\left( n-M \right) \right] ^\text{T}.
\end{equation}
And for $2\leq  r\leq  R-1$, we have 
\begin{equation}\label{eq9}
    \boldsymbol{u}_{r}=\sigma \left( \boldsymbol{W}_{r}\boldsymbol{u}_{r-1}+\boldsymbol{b}_{r} \right),
\end{equation}
where $\sigma(\cdot )$ is the activation function of the hidden layers. 
To output a full range of values, the output layer is a linear layer with no activation function.
So 
\begin{equation}
    \begin{split}
        \boldsymbol{u}_R&=\boldsymbol{W}_R\left( \boldsymbol{c}\left( n \right) \right) \boldsymbol{u}_{R-1}+\boldsymbol{b}_R\left( \boldsymbol{c}\left( n \right) \right) +\left[ u_\text{I}\left( n \right) ,u_\text{Q}\left( n \right) \right] ^\text{T}\\
        &=\left[ s_\text{I}\left( n \right) ,s_\text{Q}\left( n \right) \right] ^\text{T},
    \end{split}
\end{equation}
where $\boldsymbol{c}\left( n \right) =\left[ c_1\left( n \right) ,c_2\left( n \right) ,\ldots  \right]$ is the vector representing current operating states, 
$\boldsymbol{W}_R$ and $\boldsymbol{b}_R$ are the output layer's weights and biases controlled by $\boldsymbol{c}\left( n \right)$. 
There, $\boldsymbol{u}_{R-1}$ can be interpreted as the desired basis functions generated by the hidden layers, 
$\boldsymbol{W}_R$ and $\boldsymbol{b}_R$ can be regarded as the coefficients of these basis functions, which can be dynamically changed by operating states. 

\subsection{Offline Training Method}
In general, there are 3 architectures that can be used to train DPD models, which are indirect learning architecture (ILA), direct learning architecture (DLA), and iterative learning control (ILC)\cite{b16}. 
The ILA is is slightly inferior to DLA\cite{b15} and ILC in terms of extreme performance, 
however, it is more convenient to deploy. More importantly, in dynamic DPD scenarios, even if the number of operating 
states encountered in the offline training stage is found to be insufficient, resulting in poor actual performance, it is 
convenient to use the continual learning method\cite{b17} to adjust the DPD model online when using the ILA structure. 
So we use the ILA to train DPD models. 

When training NN-based DPD models offline, data usually need to be loaded in minibatches during each epoch. And when there are multiple operating states, 
it is important to load the data of each state in parallel (i.e., each minibatch contains part of the data of each state at the same time). 
As shown in Fig.~\ref{fig3}\subref{fig3-b}, $Q$ denotes the length of the training data for each operating state, 
$y^{(i)}(n)$ and $s^{(i)}(n)$ are training input and target output, 
and the DPD model can be expressed as 
\begin{equation}
    \tilde{s}^{\left( i \right)}\left( n \right) =f_\text{DPD}\left( \boldsymbol{y}^{\left( i \right)}\left( n \right) ,\boldsymbol{c}^{\left( i \right)};\boldsymbol{\lambda } \right),
\end{equation}
where $\tilde{s}^{\left( i \right)}\left( n \right)$ is the actual output, $\boldsymbol{y}^{\left( i \right)}\left( n \right)$ 
is the vector of training inputs, $\boldsymbol{c}^{(i)}$ is the vector representing the $i$-th operating state, and $\boldsymbol{\lambda }$ denotes all the parameters of the DPD model. 
The loss function for the $i$-th state is 
\begin{equation}
    J_i\left( \boldsymbol{\lambda}\right) =\sum_{n=1}^{Q}{|s^{\left( i \right)}\left( n \right) -\tilde{s}^{\left( i \right)}\left( n \right) |^2}.
\end{equation}
When data are loaded in parallel, the training process of NN is equivalent to solving the following optimization problem, i.e., 
\begin{equation}
    \boldsymbol{\lambda}\leftarrow \text{arg}\underset{\boldsymbol{\lambda}}{\operatorname*{\operatorname*{min}}}\left\{\sum_{i}J_{i}(\boldsymbol{\lambda})\right\}.
\end{equation}
In this way, the risk of the model gradually forgetting certain states during training is minimized.

\section{Experimental Results}\label{S4}
\subsection{Experimental Setups}
The PA in the experiment is based on the MATLAB API provided by the RF WebLab platform\cite{b20}. 
The platform uses a GaN PA operating at a center frequency of 2.14 GHz, 
its MATLAB API allows the user to input sampled I/Q data and 
remotely acquire the PA output signal, and it allows the user to 
specify the RMS power of the output signal of the vector signal generator (VSG) 
to simulate scenarios in which the PA operates at different signal power levels.
IQ imbalance is introduced artificially before the signal is fed into the 
RF WebLab platform. Frequency-independent imbalance includes a 5° phase imbalance 
and a 10\% gain imbalance, and frequency-dependent imbalance is introduced 
by two 3-rd order FIR filters with different characteristics.

We validate the proposed method using MATLAB generated 5G NR signals with bandwidth of 
(20MHz, 30MHz, 40MHz) and RMS input power of (-19dBm, -23dBm, -27dBm).
Each state contains 60000 samples for training and 20000 samples for testing.

Since it was found in this experimental platform that RMS input power has a 
slightly greater impact on DPD performance than bandwidth, the operating states received by hypernetwork is formatted as: 
\begin{equation}
    \left[ \frac{\text{BW}\left( \text{MHz} \right)}{\text{BW}_\text{max}\left( \text{MHz} \right)},\frac{\text{P}\left( \text{dBm} \right)}{\text{P}_\text{max}\left( \text{dBm} \right)},\frac{\text{P}\left( \text{dBm} \right)}{\text{P}_\text{max}\left( \text{dBm} \right)} \right].
\end{equation}

We compare the proposed HN-R2TDNN with SVDEN and HG-R2TDNN, each with a memory length of 8. The SVDEN, which consists of five layers with 18-36-18-12-2 number of neurons in each layer, 
is trained within the state of 20MHz and -19dBm, which is indexed at 1. 
The R2TDNNs used by the HN-R2TDNN and HG-R2TDNN have the same neurons in each layer as the SVDEN and they are trained using mixed data of all 9 states. 
The HG-R2TDNN includes 3 additional heterogeneous inputs and the hypernetwork used by HN-R2TDNN has 3-36-28-(12*2+2) number of neurons in each layer. 
The activation functions of the R2TDNNs and the SVDEN's hidden layers are all tanh. 

\subsection{Results}
\begin{figure}
    \centering
    \subfloat[]{\label{fig4-a}
    \includegraphics[width=0.95\linewidth]{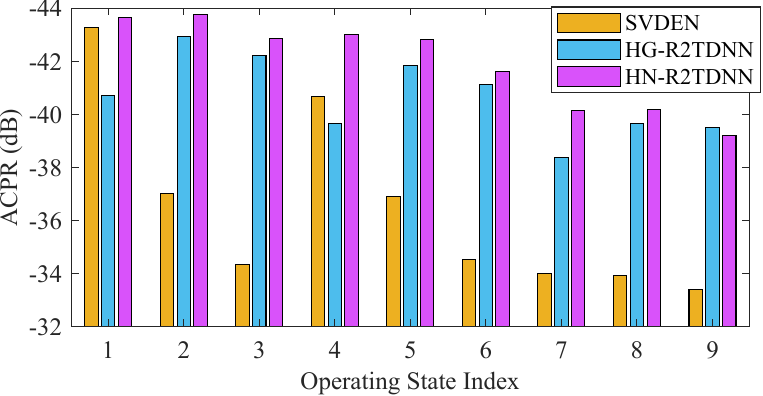}}

    \subfloat[]{\label{fig4-b}
    \includegraphics[width=0.95\linewidth]{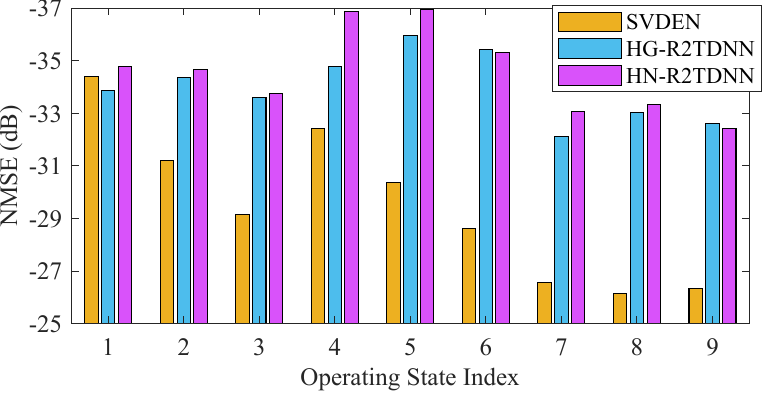}}

    \caption{The ACPR and NMSE performance of the 3 models within all 9 states. (a) ACPR. (b) NMSE.}
    \label{fig4}
\end{figure}

Fig.~\ref{fig4} shows the ACPR and NMSE performance of the 3 models within all 9 states. 
\begin{figure}
    \centering
    \includegraphics[width=0.95\linewidth]{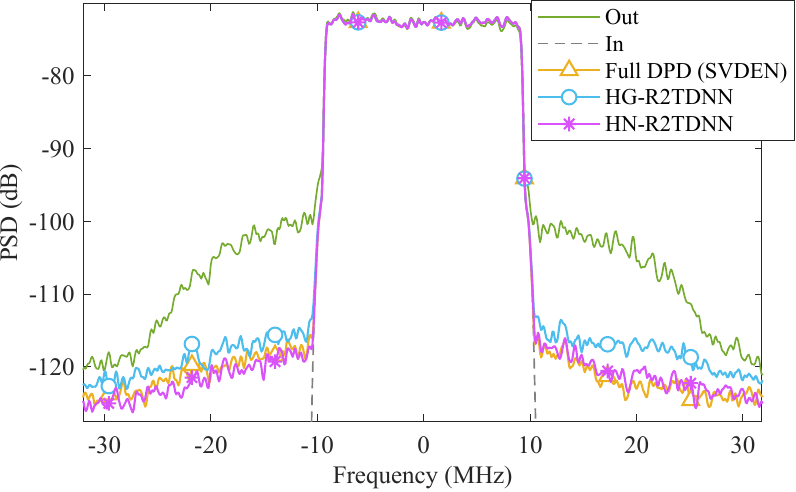}
    \caption{PSDs of the signals within operating state 1 (20MHz and -19dBm)}
    \label{fig5}
\end{figure}
It can be seen that, the proposed HN-R2TDNN can effectively perform joint DPD within all states, and 
it achieves the best overall performance, especially within states of relatively high signal power (i.e., state 1, 4 and 7). 
For example, within state 4 (30MHz and -19 dBm), the proposed HN-R2TDNN improves the ACPR by 3.36 dB and the NMSE by 2.08 dB in comparison with HG-R2TDNN. 
The SVDEN only performs well within state 1, which is its training state, and its performance degrades drastically when the bandwidth and power of the signal change. 
But even within state 1, the proposed model still performs slightly better than the SVDEN, this is because that, 
by utilizing the idea of MTL, 
the proposed HN-R2TDNN can use the information acquired within other states to help it learn better within state 1. 
The HG-R2TDNN performs better than SVDEN within most states, which indicates that the strategy of heterogeneous inputs is indeed effective when there are multiple operating states. But  
when both IQ imbalance and PA nonlinearity are considered, the characteristics of the whole system change more drastically with changes in operating states, and 
the HG-R2TDNN's performance is limited. 

Fig.~\ref{fig5} shows the power spectral density (PSD) of the signals before and after mitigation within state 1 (20MHz and -19dBm). 
It can be seen that, in terms of suppressing spectral regrowth, the proposed HN-R2TDNN is significantly better than the HG-R2TDNN, and it can keep up with or even surpass the performance of SVDEN. 

\section{Conclusion}\label{S5}
In this paper, we have proposed a new NN-based joint DPD model. 
It's based on the methodology of MLT. The hidden layers of the main NN are shared by all states, 
and the output layer's weights and biases are generated by another NN. 
Experimental results have shown that, on average, 
the proposed HN-R2TDNN has improved the ACPR by 5.46 dB and the NMSE by 5.11 dB in comparison with the SVDEN, 
which only considered stable signal states. 
And it has further improved the performance compared to the method that only considered PA nonlinearity with multiple states.

\end{document}